\documentclass[12pt,final]{iopart}

\usepackage[utf8]{inputenc}
\usepackage[T1]{fontenc}
\usepackage{xspace}
\usepackage{relsize}
\usepackage{graphicx,tikz}
\usepackage{xfrac,amssymb,bm}
\usepackage{multirow}
\usepackage[backend=biber,style=chem-angew,sorting=none]{biblatex}
\usepackage[normalem]{ulem}
\usepackage[colorlinks=true,allcolors=blue]{hyperref}

\usepackage{etoolbox}
\makeatletter
\newcommand{\mainmatter}{%
  \setcounter{footnote}{0}%
  \patchcmd{\@makefntext}{\fnsymbol}{\arabic}{}{}%
  \patchcmd{\@thefnmark}{\fnsymbol}{\arabic}{}{}%
  \def\@makefnmark{\textsuperscript{\arabic{footnote}}}%
}
\makeatother

\newcommand{\sfv}{\varphi} % {s}calar {f}ield {v}ariable
\newcommand{\sfd}{\Phi}     % {s}calar {f}ield {d}erivative (~\partial_{r}sf)
\newcommand{\sfm}{\Pi}      % {s}calar {f}ield {m}omentum   (~\partial_{t}sf)
\newcommand{\ee}{\bm{e}}
\newcommand{\DD}{\mathcal{D}}
\newcommand{\g}{\gamma}
\newcommand{\pd}{\partial}
\newcommand{\critexp}{\gamma}
\newcommand{\Lie}{\mathcal{L}}
\newcommand{\xxzero}{\mathtt{x1}}    %% xx0
\newcommand{\xxone}{\mathtt{x2}}     %% xx1
\newcommand{\xxtwo}{\mathtt{x3}}     %% xx2
\newcommand{\Nxxzero}{\mathtt{N1}}   %% Nxx0
\newcommand{\sfcol}{\texttt{SFcollapse1D}\xspace}
\newcommand{\nrpy}{\texttt{NRPy+}\xspace}
\newcommand{\nrpycol}{\texttt{NRPyCritCol}\xspace}
\newcommand{\sympy}{\texttt{SymPy}\xspace}
\newcommand{\cpp}{\mbox{C++}\xspace}
\newcommand{\python}{\texttt{Python}\xspace}
\newcommand{\jupyter}{\texttt{Jupyter}\xspace}

\newcommand{\dr}{\Delta r}
\newcommand{\dx}{\Delta x}
\newcommand{\dy}{\Delta y}
\newcommand{\dz}{\Delta z}

\newcommand{\lrpar}[1]{\left( #1 \right)}

\newcommand{\abs}[1]{\left| #1 \right|}

\newcommand{\eqref}[1]{(\ref{#1})}

\newcommand{\eq}[1]{
  \begin{equation}
    #1
  \end{equation}
}

\newcommand{\Rfd}{{}^{(4)}R} % 4-dimensional R

\newcommand{\orcid}[1]{\href{https://orcid.org/#1}{\includegraphics[height=\fontcharht\font`\B]{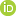}}}

\def\eadnew#1#2{\address{#2 E-mail: \mailto{#1}}}

\newcommand{\paperlongtitle}{\nrpycol \& \sfcol: an open-source, user-friendly toolkit to study critical phenomena\hspace{0.5cm}}
\newcommand{\papershorttitle}{\nrpycol \& \sfcol: a toolkit to study critical phenomena}

\begin{document}

% FIX IOPART FOOTNOTE MADNESS
\mainmatter

%% Paper title
\title[\papershorttitle]{\paperlongtitle}

%% Author list
\author{
  \mbox{Leonardo R.~Werneck$^{1,2,3,*}$~\orcid{0000-0002-4541-8553}},
  \mbox{Zachariah B.~Etienne$^{1,2,4}$~\orcid{0000-0002-6838-9185}},
  \mbox{Elcio Abdalla$^{3}$~\orcid{0000-0002-9814-1690}},
  \mbox{Bertha Cuadros-Melgar$^{5}$~\orcid{0000-0001-8111-2431}},
  \mbox{C. E. Pellicer$^{6}$~\orcid{0000-0002-6347-7813}}}
  
%% Affiliations
\address{$^{1}$ Department of Physics and Astronomy, West Virginia University, Morgantown, WV 26506, USA}
\address{$^{2}$ Center for Gravitational Waves and Cosmology, West Virginia University, Chestnut Ridge Research Building, Morgantown, WV 26505, USA}
\address{$^{3}$ Instituto de Física, Universidade de São Paulo, Caixa Postal 66318, CEP 05314-970, São Paulo, SP, Brazil}
\address{$^{4}$ Department of Physics, University of Idaho, Moscow, ID 83843, USA}
\address{$^{5}$ Escola de Engenharia de Lorena, Universidade de São Paulo, Estrada Municipal do Campinho S/N, CEP 12602-810, Lorena, SP, Brazil}
\address{$^{6}$ Escola de Ciência e Tecnologia, Universidade Federal do Rio Grande do Norte, Campus Universitário Lagoa Nova, CEP 59078-970, Natal, RN, Brazil}

%% Emails
\eadnew{wernecklr@gmail.com}{$^{*}$}

%% Abstract
\begin{abstract}

  We present a new open-source, user-friendly toolkit of two codes---\sfcol and \nrpycol---to study critical phenomena in the context of gravitational collapse. \sfcol is a C/\cpp tool designed to solve the problem of gravitational collapse of massless, spherically symmetric scalar fields with the ADM formalism in spherical-like coordinates. \nrpycol is a collection of \python modules that leverage the \nrpy infrastructure to generate a highly optimized C-code for evolving massless scalar fields within a covariant BSSN formalism. The toolkit was developed with user-friendliness and code efficiency in mind, enabling the exploration of critical phenomena with consumer-grade computers. We present a study of critical phenomena from the collapse of massless scalar fields in spherical symmetry, using only these two codes and a laptop computer.
\end{abstract}

\submitto{\CQG}

\maketitle

\section{Introduction}
\label{sec:introduction}

In 1993, Choptuik published a seminal paper announcing the discovery of critical phenomena in his studies of a massless scalar field propagating in a dynamical spacetime~\cite{choptuik1993universality}. He found a one-parameter family of solutions of the Einstein-Klein-Gordon (EKG) equations defined by the parameter $\eta$, which for example could represent the amplitude of the initial scalar field. He demonstrated that there is a critical value for this parameter, $\eta_{*}$, such that solutions with $\eta>\eta_{*}$ were comprised of spacetimes in which the scalar field collapsed and formed a black hole, while solutions with $\eta<\eta_{*}$ were characterized by full dispersion of the scalar field, resulting in flat space as the end state.

By performing several \emph{supercritical} runs (i.e. with $\eta>\eta_{*}$), Choptuik conjectured that the masses of the black holes, $M_{\rm BH}$, which were formed when $\eta\gtrsim\eta_{*}$ followed the power law
\eq{
  M_{\rm BH} \propto \abs{\eta-\eta_{*}}^{\critexp}\ ,\label{eq:Chop_BH_mass}
}
where the critical exponent $\critexp$ is \emph{universal}. The sense in which the critical exponent is said to be universal is the following: given a particular matter source, in this case a massless scalar field, the numerical value of the critical exponent is the same for all possible initial conditions. In his paper Choptuik performed runs with four different initial data families and they all resulted in the same numerical value $\critexp\approx0.37$.

Furthermore, as we improve the fine-tuning of the parameter $\eta$ and get closer to the critical solution, it possible to observe a discrete self-similarity in the solution, which repeats itself at ever-decreasing spatiotemporal scales. Choptuik found that given a value of the logarithmic proper time $\xi$ (to be defined below), the same profile of the scalar field is observed at $\xi+\Delta$, with $\Delta = 3.43$, but on a spatial scale which is $e^{\Delta}\approx 30$ times smaller than the first one. This phenomenon is known as \emph{(scale-)echoing} and $\Delta$ is the echoing period, which is also universal.

After Choptuik's discovery, many other studies were performed considering, for example, different matter sources and spacetime dimensions (see e.g.~\cite{gundlach2003critical,gundlach2007critical} for excellent reviews). In~\cite{gundlach1997understanding,martin1999all}, perturbation theory is used to show that the universal exponent is $\critexp\approx0.374$, while the echoing period has been shown to be $\Delta=3.445452402(3)$ using semi-analytical methods in~\cite{martin2003global}. Recently, the existence of the critical solution (often referred to as the Choptuik spacetime) was proven in~\cite{reiterer2019choptuik}, also using semi-analytical methods.

The power law behavior discovered by Choptuik has also been generalized in~\cite{gundlach1997understanding,hod1997fine} and now reads
\eq{
  \ln M_{\rm BH} = C + \critexp x + f(x)\ ,\label{eq:CC_mass_law}
}
where $C$ is a constant which depends on the particular initial data family and $f(x)$ is a function which is periodic in $x=\ln\abs{\eta-\eta_{*}}$ with angular frequency
\eq{
  \omega = \frac{4\pi\critexp}{\Delta}\ .
}
Given the generality of the arguments used to show the behavior \eref{eq:CC_mass_law} for the black hole mass near criticality, a quantity $\mathcal{E}$ with length dimension $\ell$ can be inferred to satisfy
\eq{
  \mathcal{E} \propto \abs{\eta-\eta_{*}}^{\ell\critexp}\ ,
}
or, more generally,
\eq{
  \ln \mathcal{E} = C + \ell\critexp \ln\abs{\eta-\eta_{*}} + f\lrpar{\ln\abs{\eta-\eta_{*}}}\ .\label{eq:CC_general_law}
}
This allows us to determine the critical exponent $\gamma$ using physical quantities other than the masses of the black holes formed in supercritical runs. In Sec.~4 we show results that illustrate this, where we have performed a series of subcritical runs and analysed the behavior of $\rho^{\rm max}_{\rm central}$, the maximum central energy density, as a function of $x$.

In his original study Choptuik solved the EKG equations using the Arnowitt-Deser-Misner (ADM) formalism~\cite{arnowitt1959dynamical}. More recently the Baumgarte-Shapiro-Shibata-Nakamura (BSSN) formalism~\cite{nakamura1987general,shibata1995evolution,baumgarte1998numerical} has been used to study critical phenomena as well~\cite{akbarian2015black,clough2016critical,baumgarte2018aspherical}, and, in particular, cases where the spherical symmetry condition is relaxed have been studied in~\cite{clough2016critical,baumgarte2018aspherical} (for similar studies that use the ADM formalism see~\cite{abrahams1993critical,choptuik2003critical}).

As is the case with many research fields that heavily rely on numerical results, critical phenomena can be a daunting subject for those who have little to no experience with solving the EKG equations numerically. When using software developed by a particular group, one often encounters the problem that they were developed ``by experts and for experts'', possessing a steep learning curve for new users. Considering the alternative track of developing one's own software, usually months (or years!) are spent writing, testing, and debugging code, which presents its own enormous barrier to entry into the field. The work we present here aims to address these issues, presenting a new efficient, highly optimized, permissively licensed, easily extensible toolkit to study critical phenomena which has been specially designed with user-friendliness in mind.

\sfcol\footnote{Available for download at: \url{https://github.com/leowerneck/SFcollapse1D}}~\cite{werneck2020aspects} is a short (just over 1800 lines of code), well-documented and open-sourced~\cite{GPLv3} \cpp code. Inspired by Choptuik's original approach, \sfcol has been specially developed to solve the EKG equations using the ADM formalism in spherical-like coordinates adopting the polar/radial gauge. This very restricted setup allowed us to greatly optimize SFcollapse1D, making it very efficient at performing the task for which it was designed. \sfcol was not implemented with extensibility in mind, but adding support for different matter sources, coordinate systems, and gauge choices is both possible and the subject of ongoing work.

\nrpycol\footnote{Available for download at: \url{https://github.com/zachetienne/nrpytutorial}}~\cite{werneck2020aspects}, on the other hand, is an open-source~\cite{BSD2c}, highly extensible, state-of-the-art numerical relativity code which uses \nrpy~\cite{ruchlin2018senr,NRPy_website} to generate highly optimized {\tt C} code kernels from Einstein-like expressions written in \python using the computer algebra package \sympy~\cite{meurer2017sympy}. Because of this, \nrpycol inherits the ability of solving the BSSN equations in the multitude of curvilinear coordinate systems that \nrpy supports (e.g., spherical-like, cylindrical-like, and Cartesian-like, just to name a few). The entire code is documented using interactive \jupyter~\cite{kluyver2016jupyter,Jupyter_website} tutorial notebooks, making the code both well-documented and intuitive to new users.

Both codes have been designed to optimize the use of memory while performing simulations, enabling the user to use consumer-grade desktop and laptop computers to study critical phenomena. As a proof of principle, all results presented in this paper have been obtained using a mid-2017 laptop computer with a 4-core Intel Core i7 CPU.

This paper is organized as follows. In Sec.~\ref{sec:basic_equations} we present a brief review of the EKG equations using the ADM formalism in the polar/radial gauge, as well as the specific form of the BSSN equations implemented in our code. In Sec.~\ref{sec:numerical_implementation} we provide an overview of the numerical methods used by our codes, explaining e.g.\ our choice of coordinate system, boundary conditions, and initial data routines. In Sec.~\ref{sec:results} we study critical phenomena of a massless scalar field using both \sfcol and \nrpycol. Finally, in Sec.~\ref{sec:conclusions} we conclude and describe plans for future research projects.

\section{Basic equations}
\label{sec:basic_equations}

Throughout this paper we adopt geometrized ($G=c=1$) units. Greek indices denote the spacetime components of indexed expressions and range from $0$ to $3$, while Latin indices denote only the spatial components and range from $1$ to $3$. We also adopt Einstein's summation convention.

We are interested in solving Einstein's field equations,
\eq{
  \Rfd_{\mu\nu} - \frac{1}{2}g_{\mu\nu}\,\Rfd = 8\pi T_{\mu\nu},
}
where $g_{\mu\nu}$ is the spacetime metric, $\Rfd_{\mu\nu}$ is the 4-dimensional Ricci tensor, \mbox{$\Rfd=g^{\mu\nu}\,\Rfd_{\mu\nu}$} is the Ricci scalar, and $T_{\mu\nu}$ is the energy-momentum tensor of a massless scalar field, $\varphi$, and is given by
\eq{
  T_{\mu\nu} = \bigl(\nabla_{\mu}\sfv\bigr)\bigl(\nabla_{\nu}\sfv\bigr) - \frac{1}{2}g_{\mu\nu}\bigl(\nabla^{\rho}\sfv\bigr)\bigl(\nabla_{\rho}\sfv\bigr),
}
where $\nabla_{\mu}$ is the covariant derivative compatible with $g_{\mu\nu}$. We cast Einstein's equations as an initial value problem by using the 3+1 formalism~\cite{arnowitt1959dynamical}, in which the spacetime metric takes the form
\eq{
  ds^{2} = g_{\mu\nu}dx^{\mu}dx^{\nu} = -\alpha^{2}dt^{2} + \gamma_{ij}\bigl(dx^{i}+\beta^{i}dt\bigr)\bigl(dx^{j}+\beta^{j}dt\bigr),\label{eq:line_element}
}
where $\alpha$ is the lapse function, $\beta^{i}$ is the shift vector, and $\gamma_{ij}$ is the spatial metric. The massless scalar field is evolved using the Klein-Gordon equation
\eq{
  \nabla^{\mu}\nabla_{\mu}\sfv = 0.\label{eq:klein_gordon}
}

Next we present two formalisms for solving the EKG system of equations numerically. In Sec.~\ref{sec:ADM_equations} we describe the fully constrained ADM formalism inspired by Choptuik's early work~\cite{choptuik1986study,choptuik1992critical,choptuik1993universality,choptuik1994critical}, and in Sec.~\ref{sec:BSSN_equations} we outline our adopted covariant BSSN formalism of Brown~\cite{brown2009covariant}.

%%%%%%%%%%
\subsection{EKG in the ADM formalism}
\label{sec:ADM_equations}

When using the ADM equations, we adopt a fully constrained evolution scheme (see e.g.~\cite{baumgarte2010numerical} for a pedagogical introduction) and adopt the polar/radial gauge, as well as spherical symmetry. The most general line element is then given by
\eq{
  ds^{2} = -\alpha(t,r)^{2}dt^{2} + a(t,r)^{2}dr^{2} + r^{2}\left(d\theta^{2} + \sin^{2}\theta d\vartheta^{2}\right),
  \label{eq:ADM_line_element}
}
where $a \equiv \sqrt{\gamma_{rr}}$ is the radial metric function. Using~\eqref{eq:ADM_line_element} the Klein-Gordon equations can be written as
\eq{
  \pd_{t}\left(\frac{a}{\alpha}\pd_{t}\sfv\right) = \frac{1}{r^{2}}\pd_{r}\left(r^{2}\frac{\alpha}{a}\pd_{r}\sfv\right)\ .\label{eq:KG}
}
Following~\cite{choptuik1993universality}, two auxiliary fields are introduced
\eq{
  \sfd\bigl(t,r\bigr)\equiv\pd_{r}\sfv\bigl(t,r\bigr)\ ;\ \sfm\bigl(t,r\bigr) \equiv \frac{a\bigl(t,r\bigr)}{\alpha\bigl(t,r\bigr)}\pd_{t}\sfv\bigl(t,r\bigr)\ ,\label{eq:SFaux}
}
enabling us to write equation~\eqref{eq:KG} as the following system of first-order partial differential equations
\eq{
  \eqalign{
    \pd_{t}\sfd &= \pd_{r}\left(\frac{\alpha}{a}\sfm\right),\cr
    \pd_{t}\sfm &= \frac{1}{r^{2}}\pd_{r}\left(r^{2}\frac{\alpha}{a}\sfd\right).\label{eq:KG_system}
  }
}
The non-trivial metric quantities are not explicitly evolved in time, but instead determined using constraint equations. The Hamiltonian constraint in the ADM formalism is given by
\eq{
  \mathcal{H} ={} R + K^2 - K_{ij} K^{ij} - 16\pi \rho = 0,\label{eq:HC_ADM}
}
where $R$ is the Ricci scalar computed from $\gamma_{ij}$, $K_{ij}$ is the extrinsic curvature, $K\equiv\gamma^{ij}K_{ij}$ is the mean curvature, and $\rho$ is the energy density. In the polar slicing, we have $K=K^{r}_{\ r}$ and \mbox{$K^{\theta}_{\ \theta} = 0 = K^{\vartheta}_{\ \vartheta}$}. Using
\eq{
  \rho = \alpha^{2}T^{tt} = \frac{\sfd^{2} + \sfm^{2}}{2a^{2}},
}
we obtain a nonlinear first-order partial differential equation (PDE) that can be solved for the radial metric,
\eq{
  \frac{\pd_{r}a}{a} + \frac{a^{2}-1}{2r} = 2\pi r \left(\sfd^{2}+\sfm^{2}\right).
  \label{eq:KG_HC}
}
Similarly, the polar slicing condition $\partial_{t}K_{\theta\theta}=0$ yields a nonlinear first-order PDE which we can solve to obtain the lapse,
\eq{
  \frac{\pd_{r}\alpha}{\alpha} - \frac{\pd_{r}a}{a} - \frac{a^{2}-1}{r} = 0.
  \label{eq:KG_polar_slicing}
}
In this coordinate system and with this gauge choice, $K^{r}_{\ r}$ can be determined using the algebraic equation
\eq{
  K^{r}_{\ r} = -4\pi r\frac{\sfd\sfm}{\alpha},
}
once we know $\bigl(\sfd,\sfm,\alpha,a\bigr)$ and therefore does not play a role in the evolution.

%%%%%%%%%%
\subsection{EKG in the BSSN formalism}
\label{sec:BSSN_equations}

We now describe the covariant formulation of the Baumgarte-Shapiro-Shibata-Nakamura (BSSN) equations~\cite{nakamura1987general,shibata1995evolution,baumgarte1998numerical} proposed by Brown~\cite{brown2009covariant}. In the BSSN formalism, we perform a conformal decomposition of the spatial metric
\eq{
  \gamma_{ij} = e^{4\phi}\bar\gamma_{ij},
}
where $\bar\gamma_{ij}$ is the conformal spatial metric, $e^{\phi}$ is the conformal factor, and $\phi$ is the conformal exponent. We also decompose the extrinsic curvature $K_{ij}$ into its trace-free and full trace parts,
\eq{
  K_{ij} = A_{ij} + \frac{1}{3}\gamma_{ij}K.
}
We introduce the conformally rescaled extrinsic curvature,
\eq{
  \bar{A}_{ij} = e^{-4\phi}A_{ij},
}
and evolve $K$ and $\bar{A}_{ij}$ instead of $K_{ij}$. Furthermore, we consider a background reference metric, $\hat\gamma_{ij}$, which is chosen to be the flat space metric in the coordinate system we are using, and define the tensor
\eq{
  \Delta\Gamma^{i}_{\ jk} \equiv \bar\Gamma^{i}_{\ jk} - \hat\Gamma^{i}_{\ jk},
}
where $\bar\Gamma^{i}_{\ jk}$ and $\hat\Gamma^{i}_{\ jk}$ are the Christoffel symbols computed from $\bar\gamma_{ij}$ and $\hat\gamma_{ij}$, respectively. As a general rule, barred quantities are associated with $\bar\gamma_{ij}$ and hatted quantities are associated with $\hat\gamma_{ij}$. We evolve the spatial vector $\bar\Lambda^{i}$, which satisfies the constraints
\eq{
  \mathcal{C}^{i} \equiv \bar\Lambda^{i} - \Delta\Gamma^{i} = 0,
}
where
\eq{
  \Delta\Gamma^{i} \equiv \bar\gamma^{jk}\Delta\Gamma^{i}_{\ jk}.
}

We adopt Brown's ``Lagrangian'' choice, in which the determinant of the conformal spatial metric is kept at its initial value throughout the numerical evolution, i.e.
\eq{
  \partial_{t}\bar\gamma = 0.
}
Defining
\eq{
  \partial_{\perp} \equiv \partial_{t} - \Lie_{\beta},
}
where $\Lie_{\beta}$ is the Lie derivative along the shift vector (see e.g. Appendix A in~\cite{baumgarte2010numerical} for an excellent review on Lie derivatives). We thus obtain the evolution equations~\cite{brown2009covariant}
\numparts
\begin{eqnarray}
  %%%%%%%%%%%%%%%%%%%%%%%%%%
  %% \bar\gamma_{ij}
  %%%%%%%%%%%%%%%%%%%%%%%%%%
  \partial_{\perp} \bar{\gamma}_{i j} &=& \frac{2}{3} \bar{\gamma}_{i j} \left (\alpha \bar{A}_{k}^{k} - \bar{D}_{k} \beta^{k}\right ) - 2 \alpha \bar{A}_{i j}, \label{eq:BSSN_gammaDD}\\
  %%%%%%%%%%%%%%%%%%%%%%%%%%
  %% \bar{A}_{ij}
  %%%%%%%%%%%%%%%%%%%%%%%%%%
  \partial_{\perp} \bar{A}_{i j} &=& - \frac{2}{3} \bar{A}_{i j} \bar{D}_{k} \beta^{k} - 2 \alpha \bar{A}_{i k} {\bar{A}^{k}}_{j} + \alpha \bar{A}_{i j} K \nonumber \\
  && + e^{-4 \phi} \bigl\{-2 \alpha \bar{D}_{i} \bar{D}_{j} \phi + 4 \alpha \bar{D}_{i} \phi \bar{D}_{j} \phi\nonumber\\
  && + 4 \bar{D}_{(i} \alpha \bar{D}_{j)} \phi - \bar{D}_{i} \bar{D}_{j} \alpha + \alpha \bar{R}_{i j} \bigl\}^{\text{TF}}, \label{eq:BSSN_ADD}\\
  %%%%%%%%%%%%%%%%%%%%%%%%%%
  %% phi
  %%%%%%%%%%%%%%%%%%%%%%%%%%
  \partial_{\perp} \phi &=& \frac{1}{6} \left(\bar{D}_{k} \beta^{k} - \alpha K \right), \label{eq:BSSN_phi}\\
  %%%%%%%%%%%%%%%%%%%%%%%%%%
  %% K
  %%%%%%%%%%%%%%%%%%%%%%%%%%
  \partial_{\perp} K &=& \frac{1}{3} \alpha K^{2} + \alpha \bar{A}_{i j} \bar{A}^{i j} - e^{-4 \phi} \left (\bar{D}_{i} \bar{D}^{i} \alpha + 2 \bar{D}^{i} \alpha \bar{D}_{i} \phi \right ), \label{eq:BSSN_K}\\
  %%%%%%%%%%%%%%%%%%%%%%%%%%
  %% \bar\Lambda^{i}
  %%%%%%%%%%%%%%%%%%%%%%%%%%
  \partial_{\perp} \bar{\Lambda}^{i} &=& \bar{\gamma}^{j k} \hat{D}_{j} \hat{D}_{k} \beta^{i} + \frac{2}{3} \Delta\Gamma^{i} \bar{D}_{j} \beta^{j} + \frac{1}{3} \bar{D}^{i} \bar{D}_{j} \beta^{j} \nonumber \\
  && - 2 \bar{A}^{i j} \left (\partial_{j} \alpha - 6 \partial_{j} \phi \right ) + 2 \alpha \bar{A}^{j k} \Delta\Gamma^{i}_{\ j k}  -\frac{4}{3} \alpha \bar{\gamma}^{i j} \partial_{j} K.\label{eq:BSSN_LambdaU}
  %%%%%%%%%%%%%%%%%%%%%%%%%%
\end{eqnarray}
\endnumparts
where $\bar{D}_{i}$ and $\hat{D}_{i}$ are covariant derivatives compatible with $\bar\gamma_{ij}$ and $\hat\gamma_{ij}$, respectively. Note that, as discussed in~\cite{baumgarte2013numerical}, terms that include the Riemann tensor associated with the flat reference metric vanish and therefore have been omitted in the above set of equations. The TF superscript denotes the trace-free part of a tensor with respect to the conformal metric so that, given an arbitrary tensor $M_{ij}$, $M^{\text{TF}}_{ij}$ is defined
\eq{
  M_{ij}^{\text{TF}} \equiv M_{ij} - \frac{1}{3}\bar\gamma_{ij}\bigl(\bar\gamma^{mn}M_{mn}\bigr).
}
Further the source terms $\rho$, $S_{i}$, $S_{ij}$, and $S$ are the energy density, momentum density, stress, and trace of the stress defined by a normal observer, respectively, and are given by
\begin{eqnarray}
  \rho &\equiv& n_{\mu}n_{\nu}T^{\mu\nu},\\
  S_{i} &\equiv& -\gamma_{i\mu}n_{\nu}T^{\mu\nu},\\
  S_{ij} &\equiv& \gamma_{i\mu}\gamma_{j\nu}T^{\mu\nu},\\
  S &\equiv& \gamma^{ij}S_{ij},
\end{eqnarray}
where
\eq{
  \gamma_{\mu\nu} \equiv g_{\mu\nu} + n_\mu n_\nu,
}
is the metric induced by $g_{\mu\nu}$ onto the spatial hypersurface, and $n^{\mu}$ is the unit vector normal to the spatial hypersurface, such that
\eq{
  n^{\mu} = \alpha^{-1}\bigl(1,-\beta^{i}\bigl)\quad\text{and}\quad n_{\mu} = \bigl(-\alpha,0,0,0\bigl).
}
The Hamiltonian constraint takes the form
\eq{
  \mathcal{H} = \frac{2}{3} K^2 - \bar{A}_{ij} \bar{A}^{ij} + e^{-4\phi} \left(\bar{R} - 8 \bar{D}^i \phi \bar{D}_i \phi - 8 \bar{D}^2 \phi\right) - 16\pi\rho= 0,
  \label{eq:BSSN_HC}
}
and the momentum constraints take the form
\eq{
  \mathcal{M}^i = e^{-4\phi} \left(\bar{D}_j \bar{A}^{ij} + 6\bar{A}^{ij}\partial_j \phi - \frac{2}{3}\bar{\gamma}^{ij}\partial_j K\right) - 8\pi S^{i}= 0.
  \label{eq:BSSN_MC}
}

Unless otherwise stated, we adopt the moving puncture gauge~\cite{campanelli2006accurate,baker2006gravitational}, in which the lapse is evolved using the ``1+log'' slicing condition~\cite{bona1995new},
\eq{
  \partial_{t}\alpha = \beta^{i}\partial_{i}\alpha-2\alpha K,\label{eq:onepluslog}
}
and the shift is evolved using the second-order covariant Gamma-driver condition suggested by Brown~\cite{brown2009covariant},
\begin{eqnarray}
  \partial_{t}\beta^{i} &=& \beta^{j}\bar{D}_{j}\beta^{i} + \frac{3}{4}B^{i},\label{eq:gammadriver1}\\
  \partial_{t}B^{i} &=& \beta^{j}\bar{D}_{j}B^{i} + \left(\partial_{t}\bar\Lambda^{i} - \beta^{j}\bar{D}_{j}\bar\Lambda^{i}\right) - \eta_{\rm S}B^{i}.\label{eq:gammadriver2}
\end{eqnarray}

When evolving the scalar field alongside the BSSN equations, we introduce the auxiliary variable
\eq{
  \sfm \equiv -n^{\mu}\nabla_{\mu}\sfv = -\frac{1}{\alpha}\left(\partial_{t}\sfv - \beta^{i}\partial_{i}\sfv\right),
}
so that~\eqref{eq:klein_gordon} can be written in terms of the BSSN variables as
\begin{eqnarray}
    \partial_{\bot}\sfv &=- \alpha\sfm,\label{eq:BSSN_sfv}\\
    \partial_{\bot}\sfm &= \alpha K\sfm - e^{-4\phi}\bar{\gamma}^{ij}\biggl[\bar{D}_{i}\bigl(\alpha\bar{D}_{j}\sfv\bigr) + 2\alpha\bar{D}_{i}\sfv\bar{D}_{j}\phi\biggr].\label{eq:BSSN_sfm}
\end{eqnarray}

\section{Numerical implementation}
\label{sec:numerical_implementation}

%%%%%%%%%%%%%%%%%%
\subsection{Solving the EKG equations: ADM formalism approach of \sfcol}
\label{sec:solving_ekg_adm}

We use finite differences to approximate all spatial derivatives appearing in the evolution and constraint equations. When evolving the fully constrained ADM system, we adopt a second-order accurate leap frog scheme in order to march the initial hypersurface forward in time. A typical time step involves advancing $\Phi$ and $\Pi$ using \eqref{eq:KG_system}, then solving the Hamiltonian constraint \eqref{eq:KG_HC} using a pointwise Newton-Raphson method in order to obtain the radial metric function $a$, and finally solving the polar slicing condition \eqref{eq:KG_polar_slicing} to obtain the lapse function $\alpha$. For the discretized versions of \eqref{eq:KG_system}, \eqref{eq:KG_HC}, and \eqref{eq:KG_polar_slicing}, \emph{cf.}\ \cite{choptuik1994critical}.

 % Coordinate system subsubsection
 \subsubsection{}
\label{sec:coord_system_adm}
\hspace*{-0.5cm}\textit{Coordinate system \& numerical grids}

\begin{figure}[htb!]
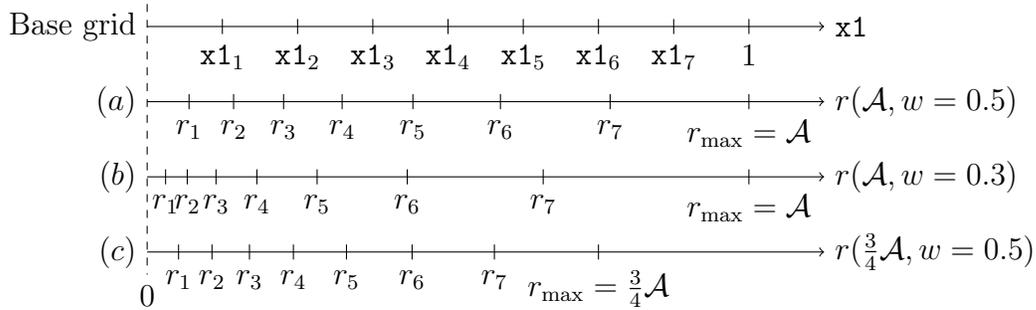

  \centering
  \tikz{
    % Draw the 0 and r_max lines
    \draw[dashed] (0.0,2.3) -- (0.0,-1.3) node [below] {$0$};
    
    % Grid #1: x-coordinate
    \draw[->] (0,2.0) node [left] {Base grid} -- (9,2.0) node [right] {$\xxzero$};
    \foreach \a in {1,...,7} {
      \draw (\a,1.9) node [below] {$\xxzero_{\a}$} -- (\a,2.1);
    }
    \draw (8,1.9) node [below] {$1$} -- (8,2.1);

    % Grid #2: r-coordinate, rmax = A, w=0.5
    \draw[->] (0.0,1.0) node [left] {$(a)$} -- (9.0,1.0) node [right] {$r(\mathcal{A},w=0.5)$};
    \foreach \a in {1,...,7} {
      \draw ({8.0*sinh(\a/8.0/0.5)/sinh(1.0/0.5)},0.9) node [below] {$r_{\a}$} -- ({8.0*sinh(\a/8.0/0.5)/sinh(1.0/0.5)},1.1);
    }
    \draw (8,0.9) node [below] {$r_{\rm max}=\mathcal{A}$} -- (8,1.1);

    % Grid #3: r-coordinate, rmax = A, w=0.3
    \draw[->] (0,0.0) node [left] {$(b)$} -- (9,0.0) node [right] {$r(\mathcal{A},w=0.3)$};
    \foreach \a in {1,...,7} {
      \draw ({8.0*sinh(\a/8.0/0.3)/sinh(1.0/0.3)},-0.1) node [below] {$r_{\a}$} -- ({8.0*sinh(\a/8.0/0.3)/sinh(1.0/0.3)},0.1);
    }
    \draw (8,-0.1) node [below] {$r_{\rm max}=\mathcal{A}$} -- (8,0.1);
    
     % Grid #4: r-coordinate, rmax = 0.75A, w = 0.5
    \draw[->] (0.0,-1.0) node [left] {$(c)$} -- (9.0,-1.0) node [right] {$r(\frac{3}{4}\mathcal{A},w=0.5)$};
    \foreach \a in {1,...,7} {
      \draw ({6.0*sinh(\a/8.0/0.5)/sinh(1.0/0.5)},-1.1) node [below] {$r_{\a}$} -- ({6.0*sinh(\a/8.0/0.5)/sinh(1.0/0.5)},-0.9);
    }
    \draw (6,-1.1) node [below] {$r_{\rm max} = \frac{3}{4}\mathcal{A}$} -- (6,-0.9);
  }
  \caption{Illustration of the coordinate system we have adopted in our runs. The top axis displays the uniformly sampled $\xxzero$-coordinate, while the other three axes display, from top to bottom, the corresponding radial coordinate $r\bigl(r_{\rm max},w\bigr)$ obtained using \eqref{eq:sinh_spherical_coordinates} for the cases $r\bigl(\mathcal{A},0.5\bigr)$, $r\bigl(\mathcal{A},0.3\bigr)$, and $r\bigl(\frac{3}{4}\mathcal{A},0.5\bigr)$, respectively.}
  \label{fig:sinh_coordinates}
\end{figure}

\noindent We adopt a coordinate system described by the variables $x^{i}=\bigl(\xxzero,\xxone,\xxtwo\bigr)$, which are uniformly sampled in the ranges $[0,1]$, $[0,\pi]$, and $[-\pi,\pi]$, respectively. The relation between these coordinates and standard spherical coordinates $\bigl(r,\theta,\vartheta\bigr)$ is given by
\begin{eqnarray}
    r &=& \mathcal{A}\frac{\sinh\bigl(\xxzero/w\bigr)}{\sinh\bigl(1/w\bigr)},\nonumber\\
    \theta &=& \xxone,\label{eq:sinh_spherical_coordinates}\\
    \vartheta &=& \xxtwo,\nonumber
\end{eqnarray}
where $\mathcal{A}=r_{\rm max}$ corresponds to the domain size and $w$ is a freely-specifiable parameter that controls how densily sampled the region near the origin is.\footnote{In \nrpy this coordinate system is known as \emph{SinhSpherical}.} In Fig.~\ref{fig:sinh_coordinates} we illustrate the radial sampling of our coordinate system.
We note that \sfcol is a 1+1 code and therefore only the coordinate $\xxzero$ in \eqref{eq:sinh_spherical_coordinates} is actually used in our implementation.

During our runs, we are mostly interested in the region near the origin. Nevertheless, the outer boundary cannot be placed too close to the center of the simulation, as spurious reflections of the outgoing waves may spoil our solution. This means that our numerical grids must be able to resolve a large range of length scales, whilst also having very high resolution near the origin, justifying our choice of the coordinate system \eqref{eq:sinh_spherical_coordinates}. By adjusting the value of $w$, the resolution near the origin can be drastically increased while keeping the outer boundary a safe distance away from the strong-field region. Typical grid structures have 320 points in the radial direction, $\mathcal{A}=16$, and $w=0.08$, resulting in resolutions of $\Delta r_{\min}\sim 1\times10^{-6}$ near the origin, while having $\Delta r_{\max} \sim 1\times10^{-1}$ near the outer boundary. The ratio $\Delta r_{\max}/\Delta r_{\min}\sim1\times10^{5}$ would require about \emph{17 levels} of traditional Berger-Oliger adaptive mesh refinement (AMR)~\cite{berger1984adaptive} by factors of 2 to resolve the same range of length scales.

Having a very high resolution near the origin greatly constrains the time steps that we are allowed to take, since we must obey that Courant-Friedrichs-Lewy (CFL) condition
\eq{
    \Delta t = \Delta r_{\rm min}\mathcal{C}_{0},
}
where $\mathcal{C}_{0}$, the CFL factor, is a number that depends on the integration scheme being used. Nevertheless, when using the ADM formalism simply decreasing the value of $w$ is enough to obtain production quality results from runs that complete in a reasonable amount of time. A typical run that uses \sfcol with the grid resolution mentioned in the previous paragraph can be performed on consumer-grade laptop computers in about 45 minutes.

 % Boundary conditions subsubsection
\subsubsection{}
\label{sec:bcs_adm}
\hspace*{-0.5cm}\textit{Boundary conditions}

\noindent \sfcol solves the EKG system in spherical-like coordinates imposing outgoing radiation boundary conditions of the form
\eq{
    \biggl[\partial_{t}\bigl(r\sfv\bigr) + \partial_{r}\bigl(r\sfv\bigr)\biggr]\bigg|_{r=r_{\rm max}} = 0.
}
The boundary condition is applied to the scalar field $\sfv$ itself, after which consistent values of the auxiliary fields $\sfd$ and $\sfm$ are computed using \eqref{eq:SFaux}. When applying boundary conditions, we use backwards finite differences whenever needed to ensure that both the spatial and time derivatives are all computed using valid data.

The PDEs that determine the radial metric and lapse functions are integrated from the origin outwards subject to the regularity conditions~\cite{choptuik1986study,choptuik1992critical,choptuik1994critical}
\eq{
    a(t,0) = 1\quad\text{and}\quad\partial_{r}\alpha(t,0) = 0.
    \label{eq:ADM_regularity}
}
In order to be able to determine the lapse using the polar slicing condition \eqref{eq:KG_polar_slicing}, we use the regularity condition to fix the value of the lapse to a constant at the origin, which is arbitrarily set to unity. This is equivalent to the statement that central observers have coordinate time equal to proper time and is a choice made by some groups (see e.g.~\cite{baumgarte2010numerical}). We choose to follow Choptuik's prescription for rescaling the lapse~\cite{choptuik1986study}, which involves computing the quantity
\eq{
    \kappa = \min\left[\frac{a}{\alpha}\left(1-|\beta|\right)\right].
}
 Remember that for our gauge choice $\beta^{i}=0$ and, thus, we get the simplified expression $\kappa = \min\left(a/\alpha\right)$. In general, the value of $\kappa$ changes from one spatial hypersurface to the next. Since the polar slicing condition is homogeneous in $\alpha$, multiplying it by a constant means that the condition is still satisfied. The particular form of $\kappa$ is chosen so that no signals propagate faster than the speed of light on our numerical grids and so that the spacetime is asymptotically flat ($\lim_{r\to\infty}\alpha\to1$). We refer the reader to Ch.~4.4 of~\cite{choptuik1986study} for further details.
 
  % Initial data subsubsection
\subsubsection{}
\label{sec:id_adm}
\hspace*{-0.5cm}\textit{Initial data}
 
 \noindent When obtaining initial data, we specify the scalar field $\sfv$ at the initial hypersurface and then analytically compute $\Phi$ from it. We assume that the initial hypersurface is at a ``moment of time symmetry'', so that $\Pi=0$ initially. We then solve the Hamiltonian constraint \eqref{eq:KG_HC} in order to determine $a$ and the polar slicing condition \eqref{eq:KG_polar_slicing} in order to determine $\alpha$, imposing the regularity conditions \eqref{eq:ADM_regularity}.

%%%%%%%%%%%%%%%%%%
\subsection{Solving the EKG equations: BSSN formalism approach of \nrpycol}
\label{sec:solving_ekg_bssn}

The time-evolution of the BSSN equations \eqref{eq:BSSN_gammaDD}-\eqref{eq:BSSN_LambdaU}, \eqref{eq:onepluslog}, \eqref{eq:gammadriver1}, \eqref{eq:gammadriver2}, \eqref{eq:BSSN_sfv}, and \eqref{eq:BSSN_sfm} is more straightforward as they are written in a form that conforms to the basic pattern expected by the method of lines (MoL):
\eq{
    \partial_{t}\vec{f} = \vec{R},\label{eq:MoL}
}
where the solution and right-hand sides vectors are denoted $\vec{f}$ and $\vec{R}$, respectively. $\vec{R}$ itself depends explicitly on the set of gridfunctions comprising $\vec{f}$ and their spatial derivatives. Derivatives appearing within $\vec{R}$ are evaluated adopting the usual centered/upwinded finite-difference approach for BSSN (outlined e.g., in~\cite{ruchlin2018senr}). Once $\vec{R}$ is evaluated, the solution vector $\vec{f}$ is stepped forward in time using a standard ODE numerical integration technique, applying boundary conditions as described in Sec.~\ref{sec:bcs_bssn}.

In this paper we adopt the fourth-order accurate Runge-Kutta method (RK4) for evolving the BSSN equations forward in time via the MoL, and fourth-order accurate finite differences for approximating the spatial derivatives appearing in $\vec{R}$. Many other MoL ODE integration options are available in \nrpy, as well as the ability to finite difference spatial derivatives to arbitrary order. Since \nrpycol is based in \nrpy, it inherits these features automatically.

% Coordinate system subsubsection
 \subsubsection{}
\label{sec:coord_system_bssn}
\hspace*{-0.5cm}\textit{Coordinate system \& numerical grids}

\noindent \nrpycol adopts the same coordinate system as \sfcol, which is described in Sec.~\ref{sec:coord_system_adm}. There are several reasons, however, that make runs much more time consuming at very high resolutions when using \nrpycol. \nrpy is a 3+1 code that does not (as of writing) support spherical symmetry, thus, requiring a minimum number of points sampling the angular directions. This can negatively impact performance as boundary conditions in the angular directions are applied needlessly. Also, and less-avoidably, the BSSN formalism contains a much larger number of gridfunctions, and the right-hand sides of their evolution equations are far more complex than the simple ADM-based system evolved within \sfcol. \nrpycol makes up for its inefficiencies with extreme flexibility and extensibility. For example, relaxing spherical symmetry, adjusting the MoL integration scheme, and increasing the finite difference order, can be performed by simply adjusting a few lines of code.

In order to dramatically speed up our runs, we have implemented a regridding infrastructure within \nrpycol that enables the grid resolution to be updated mid-run. Keeping the number of radial gridpoints fixed, the SinhSpherical (Eq.~\ref{eq:sinh_spherical_coordinates}) coordinate system offers two means for increasing the grid resolution near the origin: decreasing the value of $\mathcal{A}$ or lowering the value of $w$. Once we change the value of $\mathcal{A}$ or $w$, data from the original grid is interpolated onto a new grid, the time step is updated, and the run continues using the new grid. This technique has the advantage of adapting the resolution of the numerical grids to be well suited to the length scales associated with the solution.

We next provide a concrete example of our regridding algorithm. Let us assume that we have started a run with $\Nxxzero$ points sampling the $\xxzero$ coordinate, and with $r_{\rm max}=\mathcal{A}$ and $w=0.5$. As shown in the top part of Fig.~\ref{fig:sinh_coordinates}, SinhSpherical coordinates uniformly sample the coordinate $\xxzero$ in the interval $[0,1]$, with grid $(a)$ illustrating the corresponding radial coordinate $r$ obtained using \eqref{eq:sinh_spherical_coordinates}. Let us now consider two different possible regridding options: changing the value of $w$ so that $w_{\rm new}=0.3$ (case I) or changing the value of $\mathcal{A}$ so that $\mathcal{A}_{\rm new} = \frac{3}{4}\mathcal{A}$ (case II). By looking at Fig.~\ref{fig:sinh_coordinates}, case I/II corresponds to changing grid $(a)$ to grid $(b)$/$(c)$, respectively. Our intention is to have the same base grid after regridding, with $\xxzero \in [0,1]$ and $\Nxxzero_{\rm new} = \Nxxzero$. In order to do this, we compute $r_{\rm new} = r\bigl(\mathcal{A}_{\rm new},w\bigr)$ or $r_{\rm new} = r\bigl(\mathcal{A},w_{\rm new}\bigr)$ and then compute $\xxzero_{\rm new} = \xxzero\bigl(r_{\rm new},\mathcal{A},w\bigr)$ by inverting \eqref{eq:sinh_spherical_coordinates}. Armed with these values, we now have all the information needed to interpolate data onto the new grid ($\xxzero_{\rm new}$) from the original grid ($\xxzero$). It is important to note that numerical relativity provides only approximate outer boundary conditions, so when $\mathcal{A}$ is decreased, the physical domain becomes smaller. However, as the remaining duration of our simulation decreases with $\mathcal{A}$, spurious reflections arising from the approximate outer boundary conditions never come into causal contact with the physical region of interest (i.e., the central region of the numerical grid).

We now describe the regridding strategy we have used to obtain the results of Sec.~\ref{sec:results}. We begin our runs with a numerical grid containing 320 points in the radial direction, $\mathcal{A}=64$, and $w=0.2$. Throughout the run, we perform a total of 19 regrids to resolve the critical solution. At each of the first 11 regrids, the adjustment $w\to w-0.01$ is made to increase the radial sampling around the origin. The subsequent 8 regrids gradually shrink the outer boundary using
\eq{
    \mathcal{A}\to\mathcal{A}_{\rm new},\ \mathcal{A}_{\rm new} = \bigl\{48,32,24,16,14,12,10,8\bigr\}.
}
The total number of regrids was determined empirically by analysing the dynamics of the central values of the scalar field and the lapse function as we approached the critical solution, manually introducing additional regrids whenever the grid resolution was insufficient to resolve the features of the solution. To instill a deeper appreciation of this regridding approach, the ratio between the initial and final central resolutions for this particular case is \mbox{$\dr^{\rm initial}_{\rm min}/\dr^{\rm final}_{\rm min} \approx 1.6\times10^{3}$}. The final grid resolution ranges from \mbox{$\dr^{\rm final}_{\rm min} \approx 8.3\times10^{-6}$} near the origin to \mbox{$\dr^{\rm final}_{\rm max} \approx 2.8\times10^{-1}$} near the outer boundary, with ratio \mbox{$\dr^{\rm final}_{\rm max}/\dr^{\rm final}_{\rm min}\approx3.3\times10^{4}$}. About \emph{15 levels} of traditional (factor-of-two) AMR levels of refinement would be necessary to resolve the same range of length scales.

Results presented here that made use of the regrid infrastructure were completed within 10--15 minutes on a laptop computer. We emphasize that the regridding infrastructure is not essential to obtain results of this quality, but simply a feature that allows the user to save significant computational resources. We also note that the regridding capacities of \nrpycol are not automated yet, requiring manual user input in order to work efficiently.

 % Boundary conditions subsubsection
\subsubsection{}
\label{sec:bcs_bssn}
\hspace*{-0.5cm}\textit{Boundary conditions}

\noindent \nrpycol adopts cell-centered grids, and points lying near the boundaries of the grid interior depend upon finite-difference stencils that reach outside the interior. To address this, we extend our grids so that a certain number of points lie outside of the interior grid. These additional, exterior points are usually referred to as the \emph{ghost zone points}. 

Some of these ghost zone points exist beyond the outer boundary of our numerical domain ($r>r_{\rm max}$), and must be filled through application of outer boundary conditions. While Sommerfeld boundary conditions are available in \nrpy, we consider in this paper only simple quadratic extrapolation boundary conditions, pushing the outer boundary out of causal contact with the center of the grid to avoid any contamination resulting from spurious reflections.

Other ghost zone points in the grid exterior map directly back to points in the grid \emph{interior}. For example, in spherical coordinates, the value of a gridfunction $f(r,\theta,\vartheta)$ at a point in which $r < 0$ can be determined by simply identifying it to another point inside the grid, i.e.
\eq{
    f(-r,\theta,\vartheta) = f(r,\pi-\theta,\pi+\vartheta).
}
For a more detailed and general discussion of how these \emph{inner} boundary ghost zones are filled in the \nrpy infrastructure, see~\cite{ruchlin2018senr,CurviBCsNRPytutorialnotebooknbviewerlink}.

 % Rescaling of tensorial quantities subsubsection
\subsubsection{}
\label{sec:tensor_rescaling}
\hspace*{-0.5cm}\textit{Rescaling of tensorial quantities}

\noindent We factor out coordinate singularities from tensorial quantities and evolve the remaining regular, ``rescaled'' tensors following the procedure described in depth in~\cite{ruchlin2018senr} (see also~\cite{bonazzola2004constrained,shibata2004deriving,montero2012bssn,baumgarte2013numerical}). To briefly summarize, we write the reference metric as
\eq{
    \hat\gamma_{ij} = \delta_{kl}\ee^{(k)}_{i}\ee^{(l)}_{j},
}
where $\ee^{(i)}_{j}$ is the dual basis of the noncoordinate vectors $\ee^{i}_{(j)}$, where $(j)$ lists the individual basis vectors and $i$ its components with respect to the coordinate basis. The Kronecker delta $\delta_{ij}$ is constant and equal to the identity in all coordinate systems, which defines the noncoordinate vector basis. By definition the set $\left\{\ee_{(i)}\right\}$ is linearly independent and spans the tangent space at every point in the spatial hypersurface. Tensorial quantities are then rescaled using
\eq{
    T^{ab}_{\ \ cd} = \ee^{a}_{(i)}\ee^{b}_{(j)}\ee^{(k)}_{c}\ee^{(l)}_{d}t^{ij}_{\ \ kl},
}
where the tensor $t^{ij}_{\ \ kl}$ is well-defined on the entire hypersurface.

For example, consider the vector $\bar\Lambda^{i}$. Even if $\bar\Lambda^i$ is smooth in the Cartesian basis, Jacobians involved in the conversion to the spherical basis will generally contribute singular behavior at coordinate singularities. As the singular behavior is entirely encoded in the prescribed $\left\{\ee_{(i)}\right\}$ for spherical coordinates, the tensor rescaling procedure effectively ``factors out'' the singular part, so that the vector $\bar\Lambda^{i}$ in spherical coordinates is written in terms of the non-singular, rescaled vector $\bar\lambda^i$ as
\eq{
  \bar\Lambda^{i} =
  \ee^{i}_{\ (j)}\bar\lambda^{j}
  =
  \left(
  \matrix{
    \bar\lambda^{r}\cr
    \bar\lambda^{\theta}/r\cr
    \bar\lambda^{\vartheta}/\left(r\sin\theta\right)\cr
  }
  \right).
}
Thus, if $\bar\lambda^i$ is smooth in the Cartesian basis (where $\bar\lambda^i=\bar\Lambda^i$), $\bar\lambda^i$ will be smooth in the spherical basis. Motivations behind this rescaling procedure become transparent especially when evaluating numerical spatial derivatives of $\bar\Lambda^{i}$ in a singular coordinate system; the numerical implementation provided by \nrpy applies the product rule such that derivatives of $\bar\lambda^i$ (i.e., the smooth, rescaled variable) are entirely evaluated using numerical differentiation, and derivatives of the closed-form expressions $\left\{\ee_{(i)}\right\}$ are evaluated exactly using \sympy.

The same regularization/rescaling procedure used to obtain $\bar\lambda^{i}$ is also applied to $\bar{A}_{ij}$, $\beta^{i}$, and $B^{i}$. The conformal metric is written as
\eq{
  \bar\gamma_{ij} = \hat\gamma_{ij} + \epsilon_{ij},
}
where $\epsilon_{ij}$---the quantity to which the regularization/rescaling is applied (instead of $\bar\gamma_{ij}$ directly)---stores the deviation from the flat space metric $\hat\gamma_{ij}$, and is \emph{not} assumed to be small.

 % Numerical dissipation subsubsection
\subsubsection{}
\label{sec:dissipation}
\hspace*{-0.5cm}\textit{Numerical dissipation}

\noindent When solving the BSSN equations we add Kreiss-Oliger dissipation~\cite{kreiss1973methods} to the vector of right-hand sides $\vec{R}$. Defining $R_f$ to be element $f$ of $\vec{R}$, Kreiss-Oliger dissipation simply adds an operator $Q$ acting on $f$:
\eq{
    R_{f} \to R_{f} + Qf.
}
The standard form of the Kreiss-Oliger dissipation operator $Q$ is
\eq{
    Q \equiv \frac{\epsilon}{2^{2p}}h^{2p-1}\bigl(\DD_{+}\bigr)^{p}\bigl(\DD_{-}\bigr)^{p},
}
where $\epsilon$ is the dissipation strength, $2(p-1)$ is the accuracy of the finite difference discretization, $h$ is the step size, and $\DD_{\pm}$ are forward and backward finite difference operators. For a fourth-order accurate scheme in Cartesian coordinates, for which $p=3$, one would have
\eq{
    Q = \frac{\epsilon}{64}\left(
     \dx^{5}\frac{\partial^{6}}{\partial x^{6}}
    +\dy^{5}\frac{\partial^{6}}{\partial y^{6}}
    +\dz^{5}\frac{\partial^{6}}{\partial z^{6}}
    \right).
}
Extension to curvilinear coordinates involves normalizing each of the derivatives using $\ee^{i}_{\ (j)}$; in spherical coordinates, this becomes
\eq{
    Q = \frac{\epsilon}{64}\left(
     \Delta r^{5}\frac{\partial^{6}}{\partial r^{6}}
    +\frac{\Delta\theta^{5}}{r}\frac{\partial^{6}}{\partial\theta^{6}}
    +\frac{\Delta\vartheta^{5}}{r\sin\theta}\frac{\partial^{6}}{\partial\vartheta^{6}}
    \right).
}
The addition of $Qf$ acts to suppress high frequency oscillations on the grid, without altering the solution, as $Q\to0$ in the continuum limit. The numerical solution should also not be greatly distorted, so long as $\epsilon$ is kept small. We generally choose $\epsilon=0.1$, though values of up to 0.99 have been empirically demonstrated stable.

 % Initial data subsubsection
\subsubsection{}
\label{sec:id_bssn}
\hspace*{-0.5cm}\textit{Initial data}

\noindent When obtaining initial data in \nrpycol, we also assume that the initial hypersurface is at a ``moment of time symmetry'', which means that
\eq{
    \bar{A}_{ij} = K = \Lambda^{i} = \beta^{i} = B^{i} = \Pi = 0.
}
Furthermore, we assume the conformal metric to be initially flat. With these choices, the momentum constraints \eqref{eq:BSSN_MC} are automatically satisfied. The Hamiltonian constraint \eqref{eq:BSSN_HC} can be written in the simplified form
\eq{
    \hat{D}^{2}\psi = -2\pi\psi^{5}\rho = -\pi\bigl(\hat\gamma^{ij}\partial_{i}\sfv\partial_{j}\sfv\bigr)\psi,
    \label{eq:BSSN_SF_HC}
}
where $\psi = e^{\phi}$. One can then solve \eqref{eq:BSSN_SF_HC} to obtain the conformal factor $\psi$ by imposing the boundary conditions
\eq{
\partial_{r}\psi\bigr|_{r=0} = 0\quad \text{and}\quad \lim_{r\to\infty}\psi = 1,
}
which correspond to regularity at the origin and asymptotic flatness, respectively. The only remaining quantity is then the lapse function, which can be freely specified.  We choose the ``pre-collapsed'' value $\alpha = \psi^{-2}$.

\section{Results}
\label{sec:results}

We now present results obtained using both \sfcol and \nrpycol in the study of critical phenomena of a massless scalar field in spherical symmetry. Although we will present more results obtained using \sfcol, we emphasize that \nrpycol is by far the most versatile code. Fine-tuning of the critical solution with \nrpycol was performed with a single initial data sample, to validate against an \sfcol evolution of the same initial data.  %The choice of not fine-tuning the critical solution with many different initial data families using \nrpycol was made out of convenience.

We begin our discussion by illustrating the critical solution. Consider the initial scalar field profile
\eq{
    \sfv_{\rm I} = \eta \exp\left(-\frac{r^{2}}{\sigma^{2}}\right),
}
which corresponds to a spherically symmetric Gaussian centered at the origin, with $\eta$ a dimensionless parameter indicating its amplitude. We choose $\sigma=1$, so that all dimensionful quantities are given in units of $\sigma$.\footnote{Our problem is essentially scale-free, but setting $\sigma=1$ can be thought of as making the problem completely dimensionless by choosing a particular mass scale, e.g.\ $M_{\odot}=1$. One can then recover physical units by consider appropriate powers of $G$, $c$, and $M_{\odot}$, for example.} The subscript I indicates the initial data family, which we will need to distinguish later on.

For sufficiently small values of $\eta$ the scalar field behaves as a spherical wave which bounces at the origin and then propagates outwards, resulting in flat space as the end state of the run. On the other hand, sufficiently large values of $\eta$ lead to gravitational collapse of the scalar field with a black hole forming at the origin. We are then interested in obtaining the critical value $\eta_{*}$ which separates subcritical runs (full dispersion) from supercritical runs (gravitational collapse).

The search for $\eta_{*}$ starts by obtaining two values $\eta_{\rm weak}$ and $\eta_{\rm strong}$, which correspond to subcritical and supercritical runs, respectively. It is clear that $\eta_{\rm weak}$ and $\eta_{\rm strong}$ bracket $\eta_{*}$. One can then systematically refine the interval containing $\eta_{*}$ by using e.g.\ bisection. We define the fine-tuning parameter
\eq{
    \delta\eta \equiv \frac{\eta_{\rm strong}-\eta_{\rm weak}}{\eta_{\rm weak}},
}
which allows us to measure how close the amplitude $\eta$ is to the critical value $\eta_{*}$. The better the fine-tuning, the longer the scalar field follows the critical solution before eventually dispersing or collapsing. All results below have a fine-tuning of $\delta\eta\sim10^{-10}$ or better.

\begin{figure}[htb!]
  \centering
  \includegraphics{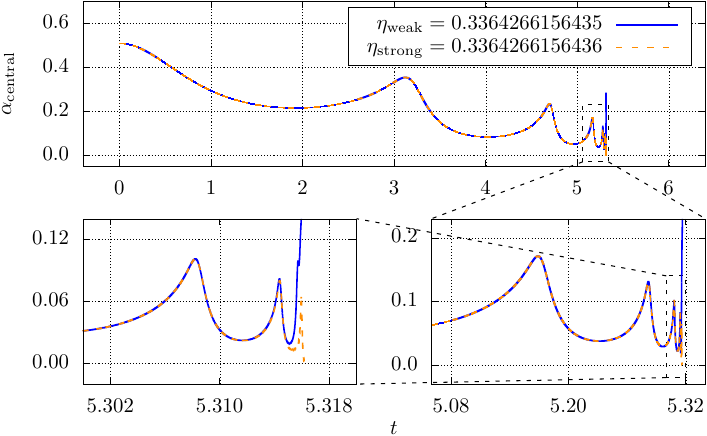}
  \caption{Central value of the lapse function as a function of coordinate time for a spherically symmetric run that adopts the ADM formalism. The top panel displays $\alpha_{\rm central}$ throughout the entire run, while the bottom two panels are zooms of increasing magnitude towards the late behavior of the lapse; the bottom-right panel is of moderate zoom, and the bottom-left panel is at a more extreme zoom. We display a subcritical (blue, solid) solution with $\eta_{\rm weak}$ and a supercritical (orange, dashed) solution with $\eta_{\rm strong}$, which bracket the critical solution for which $\eta=\eta_{*}$.}
  \label{fig:critical_lapse_ADM}
\end{figure}

\begin{figure}[htb!]
  \centering
  \includegraphics{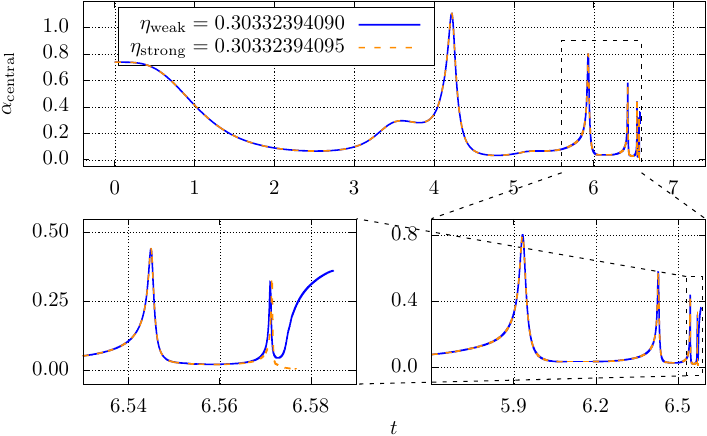}
  \caption{Same as Fig.~\ref{fig:critical_lapse_ADM} but for the BSSN formalism.}
  \label{fig:critical_alpha_BSSN}
\end{figure}

While the scalar field follows the critical solution, its profile repeats itself on ever smaller spatiotemporal scales, a phenomenon which is known as \emph{echoing}. This echoing manifests itself in other quantities as well, for example the central value of the lapse function,
\eq{
    \alpha_{\rm central} \equiv \alpha\bigr|_{r=0}.
}
In Figs.~\ref{fig:critical_lapse_ADM} and~\ref{fig:critical_alpha_BSSN} we show $\alpha_{\rm central}$ as a function of time for runs performed using the ADM and BSSN formalisms, respectively, after a high level of fine-tuning was obtained.\footnote{Instructions for reproducing Figs.~\ref{fig:critical_lapse_ADM} and \ref{fig:central_scalarfield} can be found in~\cite{sfcol_github} as well as in Sec.~5.3.2 of the supplementary material, while instructions for reproducing Fig.~\ref{fig:critical_alpha_BSSN} and the corresponding BSSN version of Fig.~\ref{fig:central_scalarfield} can be found in~\cite{ScalarFieldCollapseNRPytutorialnotebooknbviewerlink,nrpycol_original_source_code}.} Note that the difference in the overall behavior of $\alpha_{\rm central}$ between the two figures is due to the different gauge choices we made when evolving the EKG equations using the different formalisms, as described in Sec.~\ref{sec:basic_equations}.

The self-similar behavior of the central value of the scalar field is better illustrated in terms of the logarithmic proper time variable
\eq{
    \xi \equiv -\ln\bigl|\tau_{*}-\tau\bigr|,
}
where
\eq{
    \tau(t) = \int_{0}^{t}\alpha(t^{\prime})dt^{\prime},
}
is the proper time and $\tau_{*}$ is the accumulation time, after which the solution no longer follows the critical one. As we approach the accumulation time, the oscillations associated with the critical solution ``pile up'', as can be seen in Figs.~\ref{fig:critical_lapse_ADM} and \ref{fig:critical_alpha_BSSN}. This can also be observed in the first two panels of Fig.~\ref{fig:central_scalarfield}, where we show $\sfv_{\rm central}$ from a subcritical run performed with \sfcol as a function of $t$, $\tau$, and $\xi$. The same behaviour of $\sfv_{\rm central}$ is observed in runs using \nrpycol. We note that as we approach the critical solution, the oscillations of the scalar fields at the origin have a fixed amplitude of 0.61 (in our units and conventions), a behavior also observed by other groups (see e.g.~\cite{choptuik1993universality,akbarian2015black,baumgarte2018aspherical}). The same amplitude is found for all initial data families studied in this paper and the value 0.61 is also independent of the particular value of $\eta_{*}$.

\begin{figure}[htb!]
  \centering
  \includegraphics{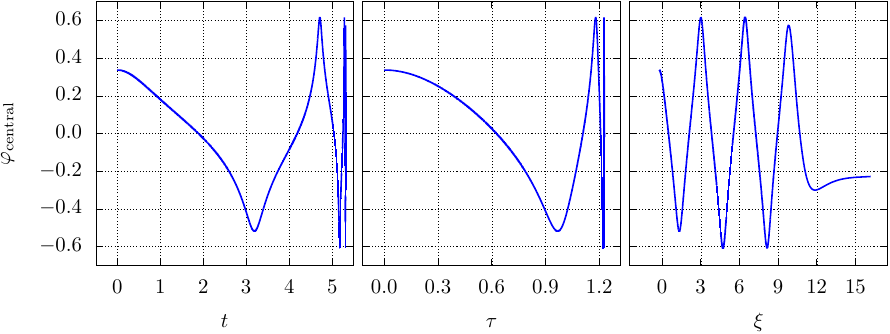}
  \caption{Central value of the scalar field near the critical solution as a function of coordinate time $t$ (left), proper time $\tau$ (center), and logarithmic proper time $\xi\equiv-\ln\left|\tau_{*}-\tau\right|$ (right). The accumulation time, $\tau_{*}$, was determined as the average value obtained from \eqref{eq:tau_star} after using all possible pairs of zero crossing proper times $\bigl(\tau_{n},\tau_{n+1}\bigr)$ for which the scalar field was already in the critical regime. We obtained $\Delta = 3.43(4)$ for the average oscillation period in logarithmic proper time, in agreement with the literature.}
  \label{fig:central_scalarfield}
\end{figure}

We follow~\cite{baumgarte2018aspherical} and consider two pairs of consecutive zero crossings of $\sfv_{\rm central}$, $\bigl(\tau_{n},\tau_{n+1}\bigr)$ and $\bigl(\tau_{m},\tau_{m+1}\bigr)$, and assume that the period of oscillation, $\Delta$, in the logarithmic proper time is constant and equal to
\eq{
  \Delta = 2\ln\left(\frac{\tau_{*}-\tau_{n}}{\tau_{*}-\tau_{n+1}}\right).
  \label{eq:universal_period_ito_crossings_times}
}
This allows us to estimate the accumulation time from the numerical data using
\eq{
  \tau_{*} = \frac{\tau_{n}\tau_{m+1}-\tau_{m}\tau_{n+1}}{\tau_{n}-\tau_{n+1}-\tau_{m}+\tau_{m+1}}. \label{eq:tau_star}
}

Next, we study the universality of the critical solution by repeating our analysis using \sfcol and the additional initial data families
\begin{eqnarray}
\sfv_{\rm II} &=& \eta r^{3}\exp\left[-\left(\frac{r-r_{0}}{\sigma}\right)^{2}\right],\\
\sfv_{\rm III} &=& \eta\left\{1 - \tanh\left[\left(\frac{r-r_{0}}{\sigma}\right)^{2}\right]\right\}.
\end{eqnarray}

We start by fine-tuning the critical parameter $\eta_{*}$ for all different families. We then perform a series of \emph{subcritical} runs (i.e. with $\eta<\eta_{*}$) and keep track of the maximum central energy density, $\rho^{\rm max}_{\rm central}$, which should behave as
\eq{
  \ln\bigl(\rho^{\rm max}_{\rm central}\bigr) = C - 2\g\,x + k\sin\bigl(\omega x + \phi_{\rm ph}\bigr),
  \label{eq:fit_critical_exponent}
}
where $\gamma$ is the universal exponent and
\eq{
  \omega = \frac{4\pi\gamma}{\Delta},
  \label{eq:universal_period_ito_omega}
}
is the frequency of the oscillation, with $\Delta$ the universal oscillation period. In the above expressions, $C$, $k$, and $\phi_{\rm ph}$ are family-dependent constants and $x\equiv\ln|\eta_{*}-\eta|$.

\begin{figure}[htb!]
  \centering
  \includegraphics{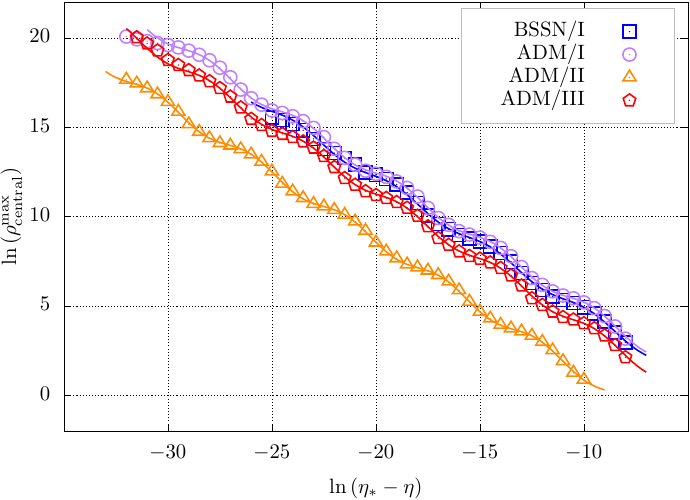}
  \caption{Maximum central density as a function of the scalar field initial amplitude $\eta$ for several subcritical runs and for different initial data families. The points represent numerical results, while the solid lines are fits obtained using~\eqref{eq:fit_critical_exponent}. The average critical exponent obtained from the fit is $\gamma=0.375$, while the average echoing period is $\Delta=3.475$. We refer the reader to Table~\ref{tab:universality_results} for a detailed summary of the results.}
  \label{fig:critical_exponent_combined}
\end{figure}

For these runs, we adopt an initial value for the amplitude of the scalar field, $\eta_{0}$, such that $x_{0}\equiv\ln|\eta_{*}-\eta_{0}|$ is set to $x_{0}=-32$ for runs that use \sfcol and $x_{0}=-25$ for runs that use \nrpycol, since the fine-tuning of the latter is smaller. We then choose a value $\Delta x=0.5$ and perform a series of runs with
\eq{
  x_{i} = x_{0} + i\cdot\Delta x,\ i=0,1,2,\ldots,N,
}
typically stopping at $x_{N} = -8$. In Fig.~\ref{fig:critical_exponent_combined} we show the results of these runs, displaying $\log\bigl(\rho^{\rm max}_{\rm central}\bigr)$ as a function of $x$.

The universal exponent $\gamma$ is determined by fitting the function \eqref{eq:fit_critical_exponent} to the numerical data. The echoing period $\Delta$ is determined in two different ways. First, using the same fitting function used to determine $\gamma$, we find the oscillation frequency $\omega$ and compute $\Delta$ using \eqref{eq:universal_period_ito_omega}. Second, we analyse the zero crossings of $\sfv_{\rm central}$ and estimate $\Delta$ using \eqref{eq:universal_period_ito_crossings_times}. In Table~\ref{tab:universality_results} we summarize our results. The uncertainties in $\gamma$ and $\Delta$ obtained using the fitting function correspond to the values for which the fit becomes visually worse to describe the numerical data. Using all possible zero crossing pairs in \eqref{eq:universal_period_ito_crossings_times}, we find the average value for $\Delta$ and its sample standard deviation, which are also presented in the table. The average value for the critical exponent is $\gamma = 0.375$, while the average value for the echoing period is $\Delta = 3.475$, both in excellent agreement with the literature.

\begin{table}[htb!]
  \centering
  \begin{tabular}{cccccccc}
    \hline
    \hline
    \multirow{2}{*}{Formalism} & \multirow{2}{*}{Family} & \multirow{2}{*}{$r_{0}$} & \multirow{2}{*}{$\sigma$} &
    \multirow{2}{*}{$\g$} & \multicolumn{2}{c}{$\Delta$} &\multirow{2}{*}{$\tau_{*}$} \\
    & & & & & \eqref{eq:universal_period_ito_omega} & \eqref{eq:universal_period_ito_crossings_times} & \\
    \hline
    BSSN & I   & -- & 1   & $0.374(3)$ & $3.51^{+0.01}_{-0.06}$ & $3.46(2)$ & $1.56275073$\\
    ADM  & I   & -- & 1   & $0.375(1)$ & $3.47^{+0.02}_{-0.04}$ & $3.43(4)$ & $1.22958674$\\
    ADM  & II  & 2  & 2   & $0.376(1)$ & $3.45^{+0.02}_{-0.03}$ & $3.42(4)$ & $4.82081707$\\
    ADM  & III & 0  & 2.5 & $0.375(1)$ & $3.47^{+0.03}_{-0.05}$ & $3.44(6)$ & $2.59392679$\\
    \hline
    \hline
  \end{tabular}
  \caption{Results for the critical exponent, echoing period, and accumulation time for different formalisms and initial data families.}
  \label{tab:universality_results}
\end{table}

\section{Conclusions}
\label{sec:conclusions}

In this paper we have presented two new open-source, user-friendly codes to study critical phenomena on consumer-grade computers: \sfcol and \nrpycol. \sfcol was designed to reproduce the original results of critical phenomena obtained by Choptuik, and serves as a good tool for new students and researchers to reproduce the seminal results in the field. \nrpycol is a state-of-the-art 3+1 numerical relativity code for studying critical collapse, and is highly extensible and versatile.

Both codes have been designed to be fast and extremely memory efficient enabling the study of critical phenomena at the scale of consumer-grade desktop and laptop computers. Additionally we make use of highly optimized numerical grids, which both exploit symmetries and are adaptively adjusted to the lengthscales of the problem.

To demonstrate the capabilities of our codes, we have presented results of a study of critical phenomena of a massless scalar field in spherical symmetry. All of these results were obtained using a mid-2017 laptop computer with a 4-Core Intel Core i7 processor. With production-quality grids, runs with \sfcol took about 45 minutes to complete, while runs with \nrpycol required as few as 10 minutes due to the implementation of manual regridding capabilities. Automation of this regridding infrastructure will be a focus of future work.

While in this paper we have focused our attention to the rather ``canonical'' example of critical collapse in spherical symmetry, \nrpycol in particular is perfectly suited for more generic scenarios. Exploration of such scenarios will also be a focus of future work.

%% Acknowledgments
\ack

We thank T.~Baumgarte, M.~Choptuik, and I.~Ruchlin for useful discussions and suggestions. LRW was financially supported in part by the Coordenação de Aperfeiçoamento de Pessoal de Nível Superior - Brazil (CAPES) - Finance Code 001. This work was also supported by NASA Grant No. TCAN-80NSSC18K1488.  ZBE gratefully acknowledges the US NSF for financial support from award PHY-1806596; and NASA for financial support from award ISFM-80NSSC18K0538. EA wishes to thank FAPESP and CNPq (Brazil) for support.

%% References
\renewcommand{\bibname}{References}
\printbibliography
%\section*{References}
%\bibliography{paper.bib}

\end{document}